\documentclass[pre,twocolumn,preprintnumbers,amsmath,amssymb]{revtex4}

\usepackage{graphicx}% Include figure files
\usepackage{dcolumn}% Align table columns on decimal point
\usepackage{bm}% bold math
\usepackage{parskip}
\usepackage{color}
\usepackage{subfigure}
\usepackage{amssymb}
\usepackage{slashed}
\usepackage{amsmath}
\usepackage{amssymb}
\usepackage{graphicx}

\def\be{\begin{eqnarray}}
\def\ee{\end{eqnarray}}
\def\bea{\begin{eqnarray}}
\def\eea{\end{eqnarray}}

\begin{document}

\title{Mono-Photon Signals  at $e^+ e^-$ Colliders in a Simplified E$_6$SSM}

\author{Shaaban Khalil}
\email[]{skhalil@zewailcity.edu.eg}
\affiliation{Center for Fundamental Physics, Zewail City of Science and Technology, 6 October City, Giza 12588, Egypt}

\author{Stefano Moretti}
\email[]{s.moretti@soton.ac.uk}
\affiliation{\small School of Physics and Astronomy, University of Southampton,
	Southampton, SO17 1BJ, United Kingdom}
	
\author{Diana Rojas-Ciofalo}
\email[]{dianitzdr@gmail.com}
\affiliation{\small National Centre for Nuclear Research
Pasteura 7, 02-093 Warsaw, Poland}
	
\author{Harri Waltari}
\email[]{h.waltari@soton.ac.uk}
\affiliation{\small School of Physics and Astronomy, University of Southampton,
	Southampton, SO17 1BJ, United Kingdom}
\affiliation{\small Particle Physics Department, Rutherford Appleton Laboratory, Chilton, Didcot, Oxon OX11 0QX, UK}

\date{\today}

\begin{abstract}
The mono-photon signature emerging in   an E$_6$ Supersymmetric Standard Model  (E$_6$SSM) from inert higgsino Dark Matter (DM)  is analyzed at future $e^+e^-$ colliders. As the inert neutral and charged higgsinos are nearly degenerate, the inert chargino is a rather long lived particle and the charged particle associated with its decay to the inert higgsino is quite soft. We show that the pair production of inert charginos at a 500 GeV electron-positron collider with an initial or final state photon is the most promising channel for probing the inert higgsino as one DM candidate within the E$_6$SSM. We also emphasize that this signal has no chance of being observed at the Large Hadron Collider (LHC) with higher energy and/or luminosity.  Finally, we remark that, combined with a DM signal produced in Direct Detection (DD) experiments  involving an active higgsino state as the second DM candidate, this dual evidence could point to a two-component DM version of the E$_6$SSM.
\end{abstract}

\maketitle
%

%%%%%%%%%%%%%%%%%%%%%%%%%%%%%%%%%%%%%%%%%%%%%%%%%%%%%%%%%%%%%%%%%%%%%%%%%%%%%%%%%%

The E$_6$ (or Exceptional) Supersymmetric Standard Model (E$_6$SSM) introduced in Refs.~\cite{King:2005jy}--\cite{King:2020ldn}
 is a natural framework for multi-component Dark Matter (DM) \cite{Athron:2016qqb}--\cite{Hall:2009aj}. In fact, in  Ref.~\cite{Khalil:2020syr}, we focused on a two-component  DM version of the E$_6$SSM based on an active and inert higgsino as candidates. We emphasized that they can share (at a comparable level) the contributions to the DM relic abundance and also illustrated that it is not possible to detect  the inert candidate in DD experiments searching for DM, whereas the active one is accessible therein. The aim of this letter is to show that it is possible to probe a light inert higgsino at $e^+ e^-$ machines, though, taking as an illustration the International Linear Collider (ILC) \cite{Djouadi:2007ik}, through a very clean mono-photon signal. Recall that we consider an E$_6$SSM along with a set of symmetries that lead to the following superpotential:    
\bea 
W &=& Y_{u} Q U^c H_u + Y_{d} Q D^c H_d + Y_{e} L E^c H_d + Y_{\nu} L \nu^c H_u  \nonumber\\
&+& \lambda S H_d H_u,
\label{Superpot}
\eea
where $H_u$, $H_d$ and $S$ are three families of doublet and singlet Higgs fields, respectively. Only $H_{u3}$,  $H_{d3}$ and $S_3$ get Vacuum Expectation Values (VEVs) while the other two families do not develop any and thus have no (or very suppressed) couplings with the SM fermions, so they remain (essentially) inert. 

In this case, the mass matrix for the inert neutralinos in the basis of \( \left(\tilde{h}^{0,I}_{d1}, \tilde{h}^{0,I}_{d2}, \tilde{h}^{0,I}_{u1}, \tilde{h}^{0,I}_{u2}\right)\) is given by 
{\small \begin{equation} 
m_{\tilde{\chi}^{0,I}} = \left( 
\begin{array}{cccc}
0 & 0 & -\frac{1}{\sqrt{2}}v_s\lambda_{311} & -\frac{1}{\sqrt{2}}v_s\lambda_{312} \\ 
0 & 0 & -\frac{1}{\sqrt{2}}v_s\lambda_{321} & -\frac{1}{\sqrt{2}}v_s\lambda_{322} \\ 
-\frac{1}{\sqrt{2}}v_s\lambda_{311} & -\frac{1}{\sqrt{2}}v_s\lambda_{312} & 0 & 0 \\ 
-\frac{1}{\sqrt{2}}v_s\lambda_{321} & -\frac{1}{\sqrt{2}}v_s\lambda_{322} & 0 & 0 
\end{array} 
\right). 
\end{equation} }

Furthermore, the mass matrix for the inert charginos in the basis of \( \left(\tilde{\chi}^{-}_{1}, \tilde{\chi}^{-}_{2}\right)\), where $\tilde{\chi}_{1(2)}=\left(\tilde{h}^{-}_{d(u)_{1}}, \tilde{h}^{-}_{d(u)_{2}}\right)$, is given by 
\begin{equation} 
m_{\tilde{\chi}^{\pm}} = \left( 
\begin{array}{cc}
-\frac{1}{\sqrt{2}}v_s\lambda_{311} & -\frac{1}{\sqrt{2}}v_s\lambda_{312} \\ 
-\frac{1}{\sqrt{2}}v_s\lambda_{321} & -\frac{1}{\sqrt{2}}v_s\lambda_{322} 
\end{array} 
\right). 
\end{equation} 

At tree-level the inert charginos and neutralinos are degenerate. Loop corrections to the mass matrix will make the chargino slightly heavier, though, with a mass splitting $m_{\tilde{h}^{\pm,I}}-m_{\tilde{h}^{0,I}}< 1$~GeV. The important thing here is that the charged particles originating from the decay of the chargino will be soft and hence will usually not be identified as leptons or jets.

In \cite{Khalil:2020syr} we considered cases where both DM components contributed a reasonably comparable share to the total relic density. In that case there are good prospects of discovering the active sector DM component through future DD experiments, but getting an unambiguous signal of the inert sector is difficult. Also, it was shown that Indirect Detection (ID) experiments were insensitive to either DM candidate.

We now look at what are the cases where we could get a signal from both DM components by relaxing the requirement that both components need to provide a substantial contribution to the relic density since, after all, these would not be extractable from such a primordial signal of DM. As the inert sector seems to be beyond the reach of DD and ID experiments, we turn to collider searches for the inert sector and rely on DD for the active sector. In order to establish the existence of two DM components, we  will also  need to show that there is enough sensitivity to the DM particle masses so that one can firmly state that there are two DM components present (of different mass).

We build our model files with \textsc{Sarah v4.14.1} \cite{Staub:2008uz,Staub:2013tta}, we compute the spectrum with \textsc{SPheno v4.0.3} \cite{Porod:2003um,Porod:2011nf} and the collider analyses are performed with \textsc{MadGraph5 v2.6.5} \cite{Alwall:2011uj}, \textsc{Pythia8} \cite{Sjostrand:2014zea}, \textsc{Delphes 3.4.2} \cite{deFavereau:2013fsa} and \textsc{MadAnalysis5 v1.8} \cite{Conte:2012fm}.

We first study the sensitivity of current LHC analyses to the possible signatures arising from the inert sector of the model using the Public Analysis Database (PAD) \cite{Dumont:2014tja} and the recasting feature of \textsc{MadAnalysis} \cite{Conte:2014zja}. The best sensitivity was achieved for
\begin{equation}
 pp\rightarrow \tilde{h}^{0,I}\tilde{h}^{\pm,I}j
\end{equation}
in the CMS multi-jet+$\slashed{E}_{T}$ analysis \cite{Sirunyan:2019ctn,recastcmssus19006}. The most sensitive signal regions were those with low jet multiplicity and no $b$-jets (bins 1--10 in \cite{Sirunyan:2019ctn}) but, even for optimistic Benchmark Points (BPs), we could exclude these only with $70\%$ Confidence Level (CL). The best exclusion power was associated with a deficit in the data compared to SM expectations. As the systematic errors were larger than the statistical ones, the expected exclusion power with $\sqrt{s}=13$~TeV and full integrated luminosity for Run 3 of the LHC is not much better.

We then turn to possible future hadron colliders. We generated events at $\sqrt{s}=14$ and $27$~TeV for the signal and background normalizing the backgrounds with those estimated by the CMS collaboration. When looking bin by bin, the cross sections for $Z\rightarrow\nu\overline{\nu}$ + jets were growing slower than those of the signal but the lost lepton backgrounds grow faster than the signal with increasing energy. Hence the sensitivity of the  High-Luminosity LHC (HL-LHC) option \cite{Gianotti:2002xx,ApollinariG.:2017ojx} or
High-Energy LHC (HE-LHC) \cite{Abada:2019ono} one are not much better than that of LHC Run 3  using this type of an analysis and seeing a significant excess would require the systematic errors to be less than half a percent, which is unrealisitic to achieve.

As hadron colliders give us little hope of establishing a signature from the inert sector in the following few decades, we decided to study the scope of $e^{+}e^{-}$ colliders. The inert neutralinos have a strongly suppressed coupling to the $Z$ boson (making them very difficult to find at DD experiments), but chargino pairs can be produced via their Electro-Magnetic (EM) couplings with a reasonable rate.

The charginos are nearly degenerate with the neutralinos so, whatever charged particles they decay to, they are soft. In most cases their transverse momentum is so small that they will not reach the outer layers of the detector and thus will not be identified as leptons or be reconstructed as jets. Therefore, the most promising channel will be the mono-photon final state as $e^{+}e^{-}\rightarrow \tilde{h}^{+,I}\tilde{h}^{-,I}\gamma$ can have photons from both initial and final state radiation.

The SM background to the mono-photon channel is dominated by $e^{+}e^{-}\rightarrow Z(\rightarrow \nu\overline{\nu})\gamma$ with a smaller contribution from the similar final state with two photons, one being missed by the detector. Such a background has a characteristic shape as a function of $x_{\gamma}=E_{\gamma}/E_{\mathrm{beam}}$, where most of the events have $x_{\gamma}\in [0.8,1.0]$ \cite{Acciarri:1998hb,Abdallah:2003np}. For the inert higgsino signal the photon energy is constrained by
\begin{equation}
E_{\gamma}< \sqrt{s}-2m_{\tilde{h}^{\pm,I}} 
\end{equation}
so that, in case of a mono-photon signal from new physics, we expect an excess of events at lower values of $x_{\gamma}$.

To illustrate this behaviour we use the two BPs given in Table \ref{tb:benchmarks}. In both cases we set the rest of the superpartner spectrum so heavy that co-annihilations in DM relic density are irrelevant and no other signals can be produced at $e^+e^-$ machines. We also require $m_{H^{I}}>m_{\tilde{H}^{0,I}}+m_{\tilde{\chi}^{0}_{1}}$, so that the inert scalar can decay and will not form a third DM component \cite{Khalil:2020syr}.

\begin{table}
\begin{center}
\begin{tabular}{|l|c|c|}
\hline
 & BP1 & BP2\\
\hline
$m_{\tilde{H}^{0,I}}$ & $173.1$ & $211.2$ \\
$m_{\tilde{H}^{\pm,I}}$ & $173.4$ & $211.6$\\
$m_{\tilde{\chi}^{0}_{1}}$ & $1117$ & $1107$\\
$m_{Z^{\prime}}$ & $4250$ & $4250$\\
BR$(\tilde{H}^{\pm,I}\rightarrow \pi^{\pm}\tilde{H}^{0,I})$ & $65\%$ & $65\%$\\
BR$(\tilde{H}^{\pm,I}\rightarrow \ell^{\pm}\tilde{H}^{0,I})$ & $35\%$ & $35\%$\\
\hline
\end{tabular}
\end{center}
\caption{The parameters of the BPs used to illustrate our results. The masses are given in GeV. For both BPs, $g_{N}=0.41$ and all other superpartners are significantly heavier than the lightest higgsino.\label{tb:benchmarks}}
\end{table}

\begin{figure}
\begin{center}
\includegraphics[width=8cm,height=5cm]{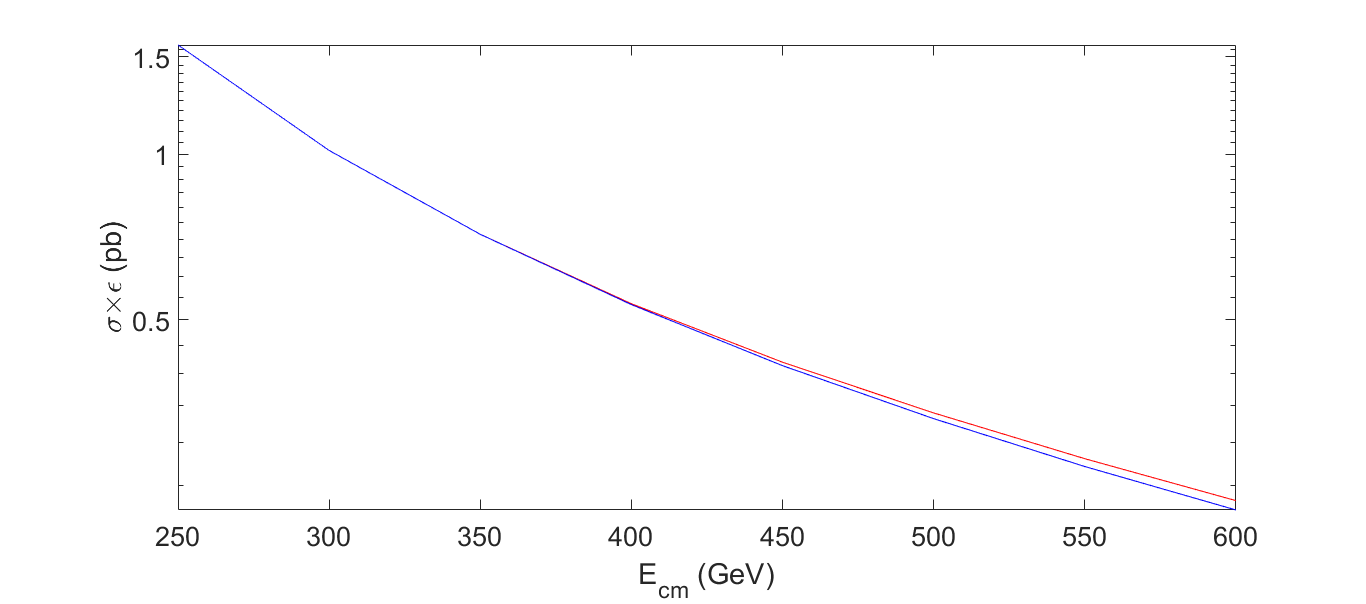}
\end{center}
\caption{The cross section times acceptance of mono-photon events with missing transverse energy satisfying $E_{\gamma}>0.1\; E_{\mathrm{beam}}$ for the SM background (blue) and SM plus the signal from BP1 (red). %The E$_6$SSM contribution gives a $3\%$ excess at $\sqrt{s}=500$~GeV.
\label{fig:xsec}}
\end{figure}

We simulate the detector with the DSiD card for Delphes \cite{Potter:2016pgp} and assume that events with $x_{\gamma}>0.1$ can be triggered. We first show the overall cross section for mono-photon events for BP1 in Figure \ref{fig:xsec}. There is an excess of a few percent once the collision energy is clearly larger than the kinematical threshold. Indeed, at about 500 GeV (a foreseen running stage on the ILC, as it would act as a factory of $t\bar t h_{\rm SM}$ events), the mono-signal should be fully established, given the precision attainable at electron-position machines in general and the ILC in particular.

\begin{figure}
\begin{center}
\includegraphics[width=8cm,height=5cm]{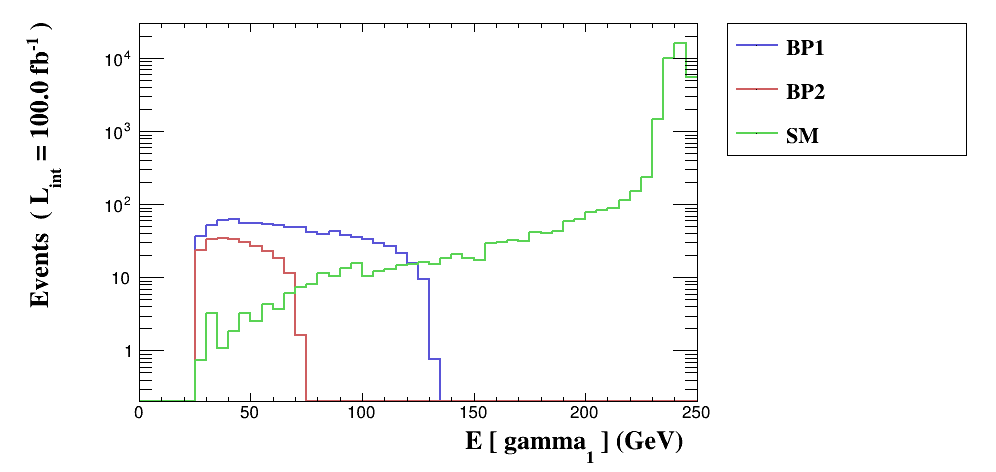}
\end{center}
\caption{The photon energy distribution for the signal benchmarks and the SM background with $\sqrt{s}=500$~GeV. 
%The signal concentrates on the lower end of the energy spectrum, while the background has photon energies close to the beam energy.
\label{fig:egamma}}
\end{figure}

In Figure \ref{fig:egamma} we show the photon energy distribution for the two BPs. The inert higgsino signal is concentrated on the lower end of the energy spectrum, where it is clearly larger than the background. 
The endpoint of the distribution even gives us a rather good estimate of the mass of the inert higgsino. 
In Table \ref{tb:yields} we show the numbers of events. With suitable cuts the signal is more than twice the background, hence, such a signal can be clearly established with moderate integrated luminosity.

\begin{table}
\begin{center}
\begin{tabular}{|l|c|c|c|}
\hline
$E_{\gamma}$ (GeV) & SM & BP1 & BP2\\
\hline
$25<E_{\gamma}<150$ & $249$ & $861$ & $236$\\
$25<E_{\gamma}<125$ & $159$ & $851$ & $236$\\
$25<E_{\gamma}<100$ & $94$ & $723$ & $236$\\
\hline
\end{tabular}
\end{center}
\caption{Photon yields in certain energy intervals for an integrated luminosity of $100$~fb$^{-1}$. The numbers for the benchmark points do not include the SM contribution.\label{tb:yields}}
\end{table}

The mono-photon signature will be sensitive as long as the collider can produce a chargino pair and a photon that exceeds the trigger threshold. We also note that the mass of the inert chargino can be estimated from the endpoint of the mono-photon excess.

\begin{figure}
	\centering
	\includegraphics[width=8cm,height=5cm]{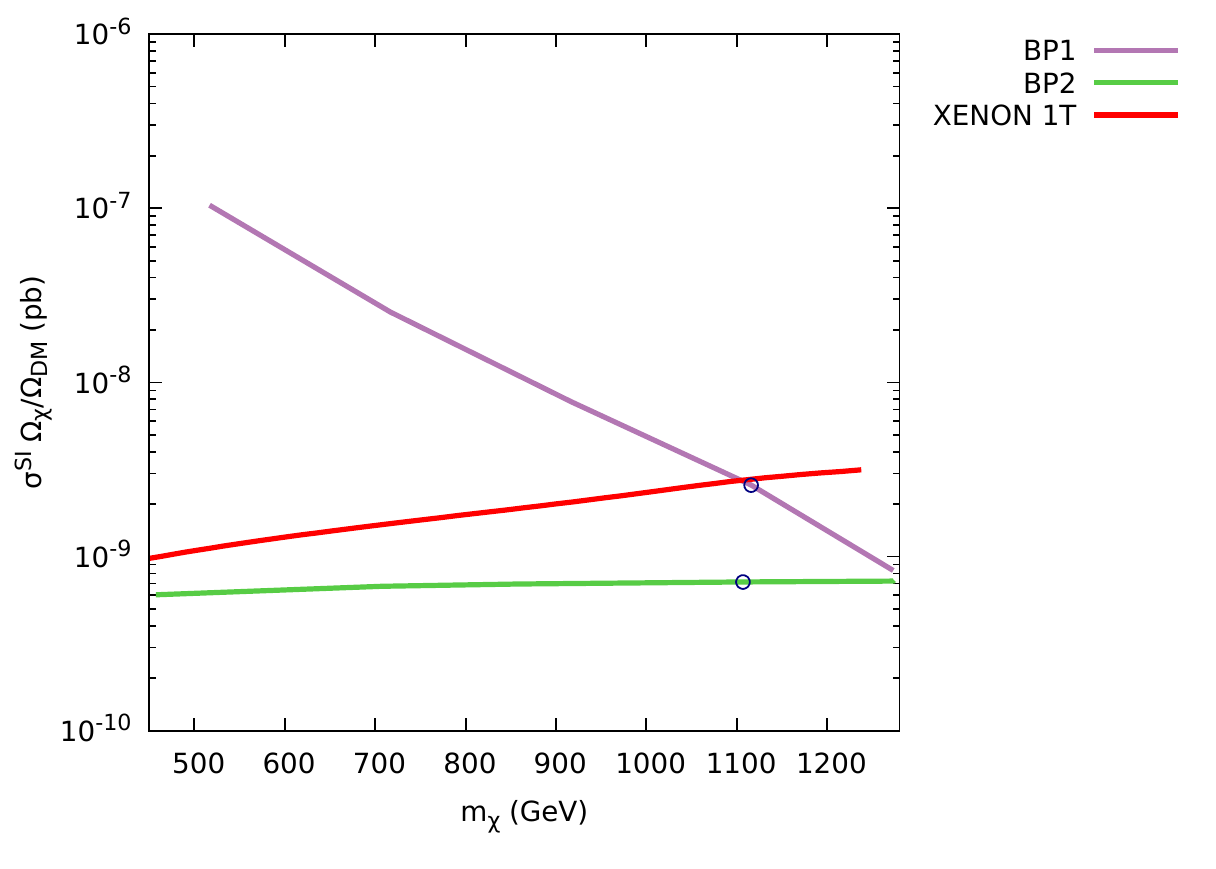}
	\caption{The Spin-Independent (SI) cross-section of the active DM component in BP1 (purple) and BP2 (green) against exclusion limits (red).}
	\label{Fig:DDactSI}
\end{figure}

\begin{figure}
	\centering
	\includegraphics[width=8cm,height=5cm]{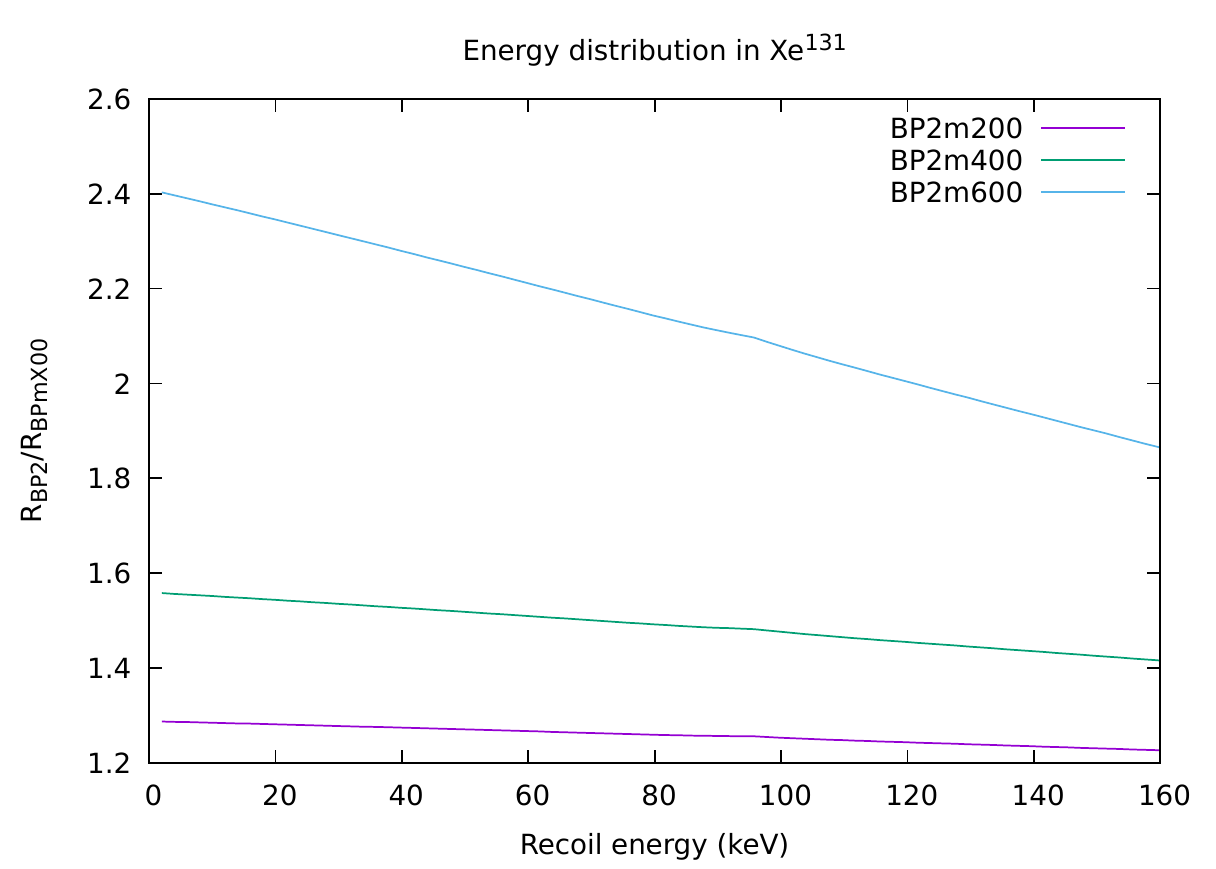}
	\caption{The recoil shape difference between different scenarios corresponding to BP2 with different masses of the active higgsino DM. }
\label{Fig:DDrecoil}
\end{figure}

We now note that mono-photon searches at the LHC are not as sensitive as those at an electron-positron collider. The reason is that, since the partonic collision energy is not fixed, the number of photons falls off with increasing energy for both the signal and background. Since there is no clear difference in the shape of the distributions, the background dominates everywhere and seeing the small excess would require sub-percent level systematic errors \cite{Baer:2014cua}. For similar reasons the sensitivity of monojet searches at the LHC is very limited \cite{Baer:2014cua,Han:2013usa}.

Furthermore, we also point out that, due to the small mass splitting between the neutralino and the chargino, the chargino is rather long-lived and could lead to disappearing track signatures. Current searches exclude radiatively split higgsinos up to $150$~GeV \cite{ATLAS:2017atz}. Hence, a disappearing track signature could further help in understanding the origin of the mono-photon signal. We shall leave the study of disappearing tracks for future work. 

Before closing, we confirm that both BPs can be accessed via  active higgsino signals in DD experiments, as shown in Figure~\ref{Fig:DDactSI} for the case of XENON 1T (wherein the circles identify their location), which can also fit the DM mass through the recoil spectra of the medium, as shown in Figure~\ref{Fig:DDrecoil} (see Ref.~\cite{Khalil:2020syr} for further details).

In summary, we have proven the sensitivity of a future $e^+e^-$ machine running at around 500 GeV (e.g., the ILC at  the $t\bar t h_{\rm SM}$ threshold) to the presence of inert higgsino signals stemming from the E$_6$SSM  in the form of mono-photon signals, with a characteristic energy spectrum dictated by the difference in mass between the parent chargino and the inert DM candidate (the lightest higgsino) of this theoretical construct. This evidence can be further complemented by the discovery in  DD experiments of another DM candidate of the E$_6$SSM, the lighest active higgsino, through the recoil spectra of the nuclei of the medium involved (e.g., in XENON 1T). As in the case of both signals one can fit the two DM masses to the relevant differential distributions, so long that these are significantly different (like in two BPs considered here), then one can point to circumstantial evidence of a two-component DM structure of a non-minimal model of Supersymmetry with origin in string theory \cite{Book}.\\

\noindent
{\bf Acknowledgments} 
The work of SK was partially supported by the STDF project 37272. SM  is  financed  in  part  through  the  NExT Institute  and  the  STFC  consolidated  Grant  No.   ST/L000296/1.
HW acknowledges financial support from the Finnish Academy of Sciences and Letters and STFC Rutherford International Fellowship (funded through the MSCA-COFUND-FP Grant No.  665593). DR-C is supported by the National Science Centre (Poland) under the research Grant No. 2017/26/E/ST2/00470.  The authors acknowledge the use of the IRIDIS High Performance Computing Facility, and associated support services at the University of Southampton, in the completion of this work.

\end{document}